%% file: main.tex
\documentclass[twocolumn]{article}
\input{Preamble}

\begin{document}

\title{Cavity-altered superconductivity}

\author[1,$\P$]{Itai Keren\thanks{ik2561@columbia.edu}}
\author[1,$\P$]{Tatiana A. Webb\thanks{tw2864@columbia.edu}}
\author[1,$\P$]{Shuai Zhang\thanks{szhangphysics@gmail.com}}
\author[1]{Jikai Xu}
\author[1]{Dihao Sun}
\author[1]{Brian S. Y. Kim}
\author[2,3]{Dongbin Shin}
\author[1]{Songtian S. Zhang}
\author[1]{Junhe Zhang}
\author[1]{Giancarlo Pereira}
\author[4,5]{Juntao Yao}
\author[1,2]{Takuya Okugawa}
\author[2]{Marios H. Michael}
\author[2]{Emil Vi\~nas Bostr\"om}
\author[6]{James H. Edgar}
\author[7]{Stuart Wolf}
\author[7]{Matthew Julian}
\author[7]{Rohit P. Prasankumar}
\author[8]{Kazuya Miyagawa}
\author[9,10,8]{Kazushi Kanoda}
\author[4]{Genda Gu}
\author[11]{Matthew Cothrine}
\author[11]{David Mandrus}
\author[2]{Michele Buzzi}
\author[2,12]{Andrea Cavalleri}
\author[1]{Cory R. Dean}
\author[2,13]{Dante M. Kennes}
\author[1,14]{Andrew J. Millis}
\author[4,15]{Qiang Li}
\author[16,2]{Michael A. Sentef}
\author[2,17]{Angel Rubio}
\author[1,4]{Abhay N. Pasupathy}
\author[1]{D. N. Basov \thanks{{db3056@columbia.edu}}}

\renewcommand{\thefootnote}{\fnsymbol{footnote}}

\affil[1]{Department of Physics, Columbia University, New York, NY 10027, USA}

\affil[2]{Max Planck Institute for the Structure and Dynamics of Matter, Luruper Chaussee 149, 22761 Hamburg, Germany}

\affil[3]{Department of Physics and Photon Science, Gwangju Institute of Science and Technology (GIST),
Gwangju 61005, Republic of Korea}

\affil[4]{Condensed Matter
Physics and Materials Science Division, Brookhaven National Laboratory, Upton, NY 11973, USA}

\affil[5]{Department of Materials Science and Chemical Engineering, Stony Brook University, Stony Brook, NY 11794-3800, USA}

\affil[6]{Tim Taylor Department of Chemical Engineering, Kansas State University, Manhattan, KS 66506, USA}

\affil[7]{Deep Science Fund, Intellectual Ventures, Bellevue, WA 98007, USA}

\affil[8]{Department of Applied Physics, The University of Tokyo, Bunkyo, Tokyo 113-8656, Japan}

\affil[9]{Max Planck Institute for Solid State Research, Heisenbergstrasse 1,
70569 Stuttgart, Germany}

\affil[10]{Physics Institute,  University of Stuttgart, Pfaffenwaldring 57, 70569
Stuttgart, Germany}

\affil[11]{Department of Materials Science and Engineering, University of Tennessee, Knoxville, TN 37996, USA}

\affil[12]{Department of Physics, Clarendon Laboratory, University of Oxford, Oxford OX1 3PU, United Kingdom}

\affil[13]{Institut für Theorie der Statistischen Physik, RWTH Aachen, 52056 Aachen,
Germany and JARA - Fundamentals of Future Information Technology}

\affil[14]{Center for Computational Quantum Physics, The Flatiron Institute,
New York, NY, USA}

\affil[15]{Department of Physics and Astronomy, Stony Brook University, Stony Brook, NY 11794-3800, USA}

\affil[16]{Institute for Theoretical Physics and Bremen Center for Computational Materials Science,
University of Bremen, 28359 Bremen, Germany}

\affil[17]{Initiative for Computational Catalysts, The Flatiron Institute,
New York, NY 10010, USA}

\affil[$\P$]{These authors contributed equally}

\date{}
\maketitle

\textbf{\boldmath Is it feasible to alter the ground state properties of a material by engineering its electromagnetic environment?  Inspired by theoretical predictions~\cite{sentef2018cavity,ruggenthaler2018quantum,PriNarang,garcia2021manipulating,latini2021ferroelectric,schlawin2022cavity,bloch2022strongly,ebbesen2023introduction,vinas2023controlling,hubener2024quantum,Rubio_Cavity,lu2025cavity,Fausti_Theory,riolo2025tuning}, experimental realizations of such cavity-controlled properties without optical excitation are beginning to emerge~\cite{paravicini2019magneto,appugliese2022breakdown, fausti2023cavity, Thomas2021,enkner2024testing,thomas2025,enkner2025tunable}. Here, we devised and implemented a novel platform to realize cavity-altered materials. Single crystals of hyperbolic van der Waals (vdW) compounds provide a resonant electromagnetic environment with enhanced density of photonic states and prominent mode confinement~\cite{basov2016,low2017polaritons,Polariton_Panorama,herzig2024high,basov2025polaritonic}.
We interfaced hexagonal boron nitride (hBN) with the molecular superconductor $\kappa$-(BEDT-TTF)$_2$Cu[N(CN)$_2$]Br (\ET). The frequencies of infrared (IR) hyperbolic modes of hBN~\cite{basov2104,caldwell2014sub} match the IR-active carbon-carbon stretching molecular resonance of {\ET} implicated in superconductivity~\cite{Cavalleri2020}. Nano-optical data  supported by first-principles molecular Langevin dynamics simulations confirm the presence of  resonant coupling between the hBN hyperbolic cavity modes and the carbon-carbon stretching mode in {\ET}.
Meissner effect measurements via magnetic force microscopy demonstrate a strong suppression of superfluid density near the hBN/{\ET} interface. Non-resonant control heterostructures, including {\RuCl}/{\ET} and hBN/$\text{Bi}_2\text{Sr}_2\text{CaCu}_2\text{O}_{8+x}$, do not display the pronounced superfluid suppression. These observations suggest that hBN/{\ET} realizes a cavity-altered superconducting ground state. Our work highlights the potential of dark cavities devoid of external photons for engineering electronic ground state properties of complex quantum materials.
    }

Some of the most exciting properties of solids arise from strong collective interactions among electrons, spins, and the crystal lattice. 
The emergent effects born out of such strong interactions are abundant and drive the formation of  varied electronic and magnetic phases. 
At the vanguard of current interest and debate is the question whether the strong interaction of the quantum fluctuations of electromagnetic modes in photonic cavities or meta-structures with elementary excitations in solids 
can also prompt phase transitions and give rise to new quantum states of matter.
 A grand aspiration of cavity quantum materials research is to uncover fundamentally new routes for controlling  properties of matter by judiciously tailoring the quantum electromagnetic environment~\cite{sentef2018cavity,ruggenthaler2018quantum,PriNarang,garcia2021manipulating,latini2021ferroelectric,schlawin2022cavity,bloch2022strongly,ebbesen2023introduction,vinas2023controlling,hubener2024quantum,Rubio_Cavity,lu2025cavity,Fausti_Theory,riolo2025tuning}.
 Experiments with dark cavities revealed modified transport properties in the integer and fractional quantum Hall states of a 2D electron gas~\cite{paravicini2019magneto,enkner2024testing,enkner2025tunable} as well as cavity-assisted thermal control of the metal-to-insulator transition in charge density wave systems~\cite{fausti2023cavity}.
Pioneering theoretical works on cavity control of superconductivity explored dark-cavity modification of phonon-mediated pairing through an increase in electron-phonon coupling via phonon polaritons \cite{sentef2018cavity,hubener2024quantum,Rubio_Cavity}, Amperean pairing directly mediated through dark-cavity photons \cite{schlawin2019,polini2024} as well as the driven-cavity extension of the Eliashberg effect \cite{curtis2019}.
 

Here, we have chosen to focus on superconductivity as a purely electronic phase transition, devoid of lattice reconstructions and charge or spin orderings. We work with $\kappa$-(BEDT-TTF)$_2$Cu[N(CN)$_2$]Br (\ET), a widely studied layered organic salt that superconducts below the transition temperature $\Tc$=11.5~K~\cite{kanoda1991nmr,dressel1994electrodynamics} (Supplementary Information Sections 1,2). 
We harness an electromagnetic environment structured by the dipole-active phonons of a thin van der Waals (vdW) hyperbolic material, hexagonal boron nitride (hBN). Hyperbolicity occurs when the permittivity has opposite signs along different axes~\cite{basov2104, caldwell2014sub} and results in a highly enhanced photonic density of states~\cite{Poddubny2013,Sun2014,Kim:12}. 
Notably, hyperbolic modes  (HMs) of hBN overlap with the frequency of the carbon-carbon (C=C) stretching mode of {\ET} at 1470~\wavenum~\cite{ET_epsilon} and the two modes hybridize at the hBN/{\ET} interface.
Effectively, the hBN slab acts as an electromagnetic cavity (Supplementary Information Section 3) resonantly tuned to the C=C stretching mode of {\ET} that is implicated in superconductivity~\cite{Cavalleri2020}. 
In this work, we examine how superconductivity is altered at the hBN/\ET\ interface, using advanced scanning probe-based techniques adept at the analysis of electrodynamics in buried interfaces.

\begin{figure*}[t]
    \centering
    \includegraphics[]{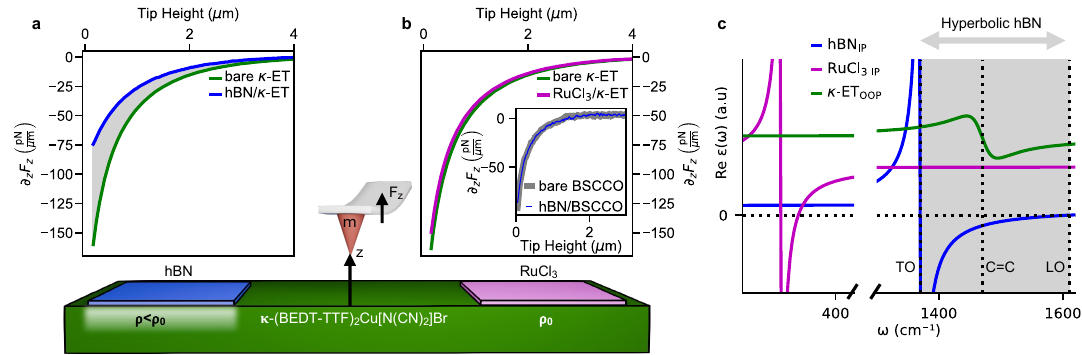} 
    \caption{ \textbf{\boldmath Electromagnetic environment alters the superfluid density at the interface between hBN and $\kappa$-(BEDT-TTF)$_2$Cu[N(CN)$_2$]Br ({\ET}). }Schematic at the bottom depicts the Meissner force $F_z$ experienced by the magnetic atomic force microscope tip above the surface of the {\ET} molecular superconductor. Exfoliated hBN and {\RuCl} microcrystals sit on the surface of a bulk {\ET} crystal (optical images in Supplementary Information Section 4). \textbf{a,b,} Magnetic force microscopy data displayed in the form of the derivative of the Meissner force $\dfdz$ as a function of tip height on bare \ET\ (green), hBN/\ET\ (blue), and {\RuCl}/{\ET} (magenta), taken at temperature 2 K. The \ET\ surface is at $z=0$, and the gray shading highlights the difference between the two curves.
    The vertical axes in panels \textbf{a} and \textbf{b} are identical. The superfluid density is significantly reduced near the hBN/{\ET} interface, but not near the {\RuCl}/{\ET} interface. Inset: same as \textbf{a}, except that the superconductor has been replaced by $\text{Bi}_2\text{Sr}_2\text{CaCu}_2\text{O}_{8+x}$ (BSCCO). The bare BSCCO curve is plotted using a wider gray line for visual clarity.
    \textbf{c,} Model real part of the out-of-plane (OOP) permittivity for {\ET}  (green) and the in-plane (IP) permittivities for hBN (blue) and {\RuCl} (magenta), based on refs.~\cite{ET_epsilon,hBN_epsilon,RuCl3_DC_epsilon}. For clarity, the permittivities have been simplified to show only the relevant modes. The C=C stretching mode of {\ET} (labeled C=C) falls within the hyperbolic region of hBN between the transverse optical (TO) and longitudinal optical (LO) frequencies (gray shading).
    }
    \label{fig:z-specs}
\end{figure*}

\section*{Meissner effect at resonant and non-resonant interfaces}

{\mainpoint Magnetic force microscopy (MFM) probes the local superfluid density via the repulsive Meissner force experienced by a magnetized tip at the end of a cantilever oscillating above the surface of a superconductor~\cite{xu1995magnetic, Luan2010,ophir2015,grebenchuk2020crossover}}.
The force results from the supercurrent screening the stray magnetic field of the tip.
Notably, the Meissner force is not affected when non-magnetic materials are inserted between the tip and the superconductor, and
the data in Fig.~\ref{fig:z-specs}(a,b) attest that MFM probes superconductors encapsulated by insulating layers. 
We measure the shift in the resonant frequency of the cantilever, which is proportional to the force gradient $\dfdz$~\cite{Hartmann1999}.
Analysis of $\dfdz$ allows the extraction of the superfluid density $\rho_0$~\cite{Luan2010} (Methods and Supplementary Information Section 5).

We examine the impact of hBN on superconductivity in a representative  heterostructure composed of a 60-nm-thick hBN microcrystal on the surface of a bulk {\ET} single crystal ($\sim$200 $\mathrm{ \mu m}$ thick). We report $\dfdz$ measured as a function of $z$, the separation between the tip and the {\ET} surface, in Fig.~\ref{fig:z-specs}a. On a region of the \ET\ crystal unobscured by hBN (green), $\dfdz(z)$ is negative  at $T=2$ K, indicating a repulsive force. The magnitude of $\dfdz(z)$ grows with reduced tip-sample surface separation. The $\dfdz(z)$ signal collected above hBN (blue) shows a  weakened $\dfdz$, highlighted by the gray shading. 
As a control experiment, we placed a 55-nm-thick {\RuCl} microcrystal on top of the same {\ET} sample.  {\RuCl} is an insulator with a static permittivity similar to that of hBN~\cite{hBN_DC_epsilon,RuCl3_DC_epsilon}. However, optical phonons of {\RuCl} all occur at much lower frequencies below 350 cm$^{-1}$~\cite{RuCl3_DC_epsilon} and therefore do not resonantly couple to the C=C stretching mode of {\ET}. Fig.~\ref{fig:z-specs}c schematically depicts the permittivities of hBN, {\ET}, and {\RuCl}. 
Comparing measurements above {\RuCl/\ET} (Fig.~\ref{fig:z-specs}b, magenta curve) and bare {\ET} (green curve - same as in Fig.~\ref{fig:z-specs}a), we observe a much weaker (less than $\sim7$\%) effect than over hBN/{\ET}. 

The stark contrast between the {\RuCl/\ET} and hBN/{\ET}  interfaces indicates that the suppression of superfluid density under hBN is likely prompted by resonant coupling across the hBN/{\ET} interface.
We remark that the {\RuCl} control experiment rules out a prominent role of charge transfer due to work function disparity~\cite{rizzo2020charge} in the suppression of the superfluid density in Fig.~\ref{fig:z-specs} (Supplementary Information Section 14) and similarly suggests that strain is unlikely to be significant.
We perform a second control measurement on a heterostructure of 8-nm hBN residing on a 28.5-nm-thick microcrystal of $\text{Bi}_2\text{Sr}_2\text{CaCu}_2\text{O}_{8+x}$ (BSCCO) with $\Tc=55$~K. Crystals of BSCCO host phonons below 650 cm$^{-1}$~\cite{kovaleva2004c}, far below the HMs in hBN. MFM data plotted in Fig.~\ref{fig:z-specs}b (inset) for hBN/BSCCO (blue) are indistinguishable from the data for bare BSCCO (dark gray - Supplementary Information Section 6).
Thus, the significant suppression of the superfluid density is specific to the resonant nature of the  hBN/{\ET} interface.

\section*{\boldmath Suppression of superfluid density at the hBN/\texorpdfstring{\ET}{ET} interface}

\begin{figure*}[ht]
    \centering
    \includegraphics[]{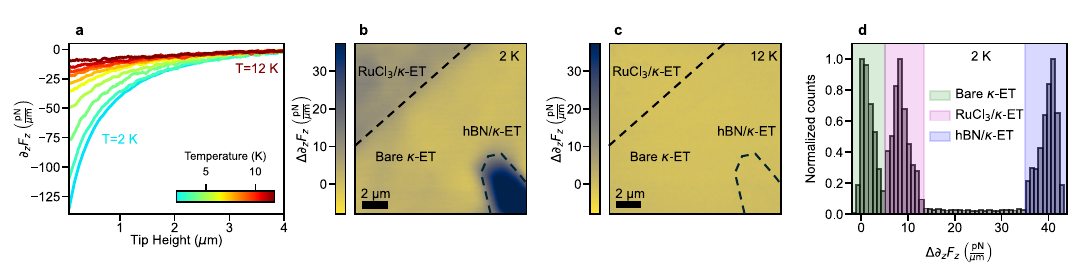}

    \caption{\textbf{Meissner effect by magnetic force microscopy nano-imaging.} 
    \textbf{a,} $\dfdz$ measured as a function of tip height at temperatures from 2~K to 12~K on bare {\ET}. The curves flatten as the temperature increases (colorscale). 
    \textbf{b, c,} Constant-height MFM images over an area with hBN/\ET, bare \ET, and \RuCl/\ET\ regions all within the same field of view, taken at a temperature of 2 K (below $\Tc$) with a tip height of 300~nm above the {\ET} surface (b) and at 12~K (above $\Tc$)  with a tip height of 150~nm above {\ET} (c). The boundaries of the areas covered by hBN and {\RuCl} are marked by black dashed lines. The false color map represents the differential Meissner signal $\Deltadfdz$.  
    \textbf{d,} Histogram of $\Deltadfdz$ values extracted from the image in panel b. Green, magenta, and blue shading highlight typical $\Deltadfdz$ values measured over bare {\ET}, over the {\RuCl}/{\ET} interface, and over the hBN/{\ET} interface respectively.
    }
    
    \label{fig:imaging}
\end{figure*}

We now validate the association of reduced $\dfdz$ (Fig.~\ref{fig:z-specs}a) with suppressed superfluid density by studying the $\dfdz$ temperature dependence. 
Over bare {\ET}, the large repulsive $\dfdz$ arises from the Meissner effect (Supplementary Information Section 5), decreases with increasing temperature (Fig.~\ref{fig:imaging}a), and vanishes near $\Tc$.
Having confirmed that Meissner repulsion is the main contributor to $\dfdz$ over bare {\ET}, we examine the effects at the heterostructure interfaces. We performed constant-height scans with the tip hovering above the {\ET} surface in a 16$\times$16 $\mu$m$^2$ region that includes both hBN and {\RuCl} (Fig.~\ref{fig:imaging}b,c). To facilitate comparison between data collected at different temperatures, we focus on the differential signal $\Deltadfdz(x,y) = \dfdz(x,y) - \dfdz|_\textrm{bare}$, where the latter term stands for $\dfdz$ over bare {\ET}. 
At 2~K (Fig.~\ref{fig:imaging}b), $\Deltadfdz \sim 40$~pN/$\mu$m over hBN, indicating suppressed superfluid density. Over {\RuCl}, $\Deltadfdz$ is visibly much smaller. More quantitatively, in Fig.~\ref{fig:imaging}d we display the histogram corresponding to the image of the $\Deltadfdz$ signals. We witness distinct distributions representing bare {\ET}, {\RuCl}/{\ET}, and hBN/{\ET}. The hBN/{\ET} peak is prominently separated from the other two, indicating the strong effect of hBN on the superfluid density of {\ET}.  Supplementary Information Sections 7 and 8 discuss additional effects, including possible scenarios for the minor difference between the MFM data registered at the {\RuCl}/{\ET} interface and the bare {\ET}.
Upon warming to 12~K, just above $\Tc$, the image contrast disappears, giving $\Deltadfdz=0$ everywhere (Fig~\ref{fig:imaging}c). The vanishing $\Deltadfdz$ near $\Tc$ confirms that the weakened $\dfdz$ signal observed at 2~K is associated with a reduction in the superfluid density.

\begin{figure}[ht]
    \centering
    \includegraphics[]{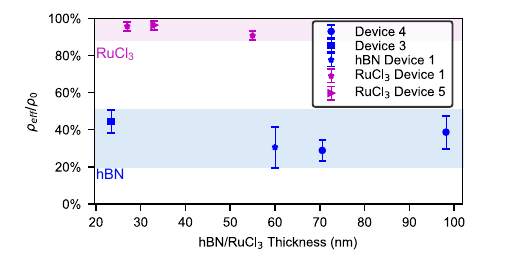}
    \caption{\textbf{Quantifying superfluid density suppression.}
    $\rho_\textrm{eff}/\rho_\textrm{0}$ (defined in the main text) shown as a function of cavity thickness. Measurements on {hBN}/{\ET} and {\RuCl}/{\ET} are shown in blue and magenta, respectively.
    Symbols distinguish different devices. 
    The superfluid density is assumed to be uniformly suppressed underneath the hBN and {\RuCl} microcrystals. Error bars correspond to the error in local superfluid density due to uncertainty in tip geometry (Supplementary Information Section 5) and exclude spatial variations of the hBN/{\ET} interface (Supplementary Information Section 8).} 
     
    \label{fig:Quantitative}
\end{figure}

We observe suppression of superfluid density under hBN in multiple hBN/{\ET}  heterostructures. The magnitude of suppression varies spatially (Supplementary Information Section 8); we focus on regions with the strongest effect.
The results are summarized in Fig.~\ref{fig:Quantitative}.
To model $\dfdz$ analytically~\cite{xu1995magnetic}, we make the grossly simplifying assumption that the suppression of the superfluid density is uniform in depth and evaluate the effective superfluid density under hBN, $\rhoeff$.  In Fig.~\ref{fig:Quantitative}, we compile MFM data for several devices with hBN cavities and {\RuCl} dielectric layers of varied thicknesses. 
We present data in the form of $\rhoeff / \rho_0$, where $\rho_0$ is the superfluid density in bare {\ET}. 
In all heterostructures, we observe at least 50\% suppression of the superfluid density prompted by the hBN cavity.
This result holds over a large variation in the hBN thickness, ranging from 25~nm to 110~nm. 
Accurate conversion between $\dfdz$ and superfluid density would require knowledge of the depth profile (Supplementary Information Section 5). 
While we expect the suppression of superfluid density to be strongest in the immediate vicinity of hBN and to decay into the {\ET} bulk, MFM measurements alone cannot differentiate among the possible depth profiles that all lead to similar $\dfdz$. Nevertheless, the magnitude of the observed suppression of $\dfdz$ implies that superconductivity in {\ET} is impacted by the cavity to a significant depth (Supplementary Information Section 9).


\section*{\boldmath Electrodynamics of the hBN/\texorpdfstring{\ET}{ET} interface}

\begin{figure*}[t]
    \centering
    \includegraphics[]{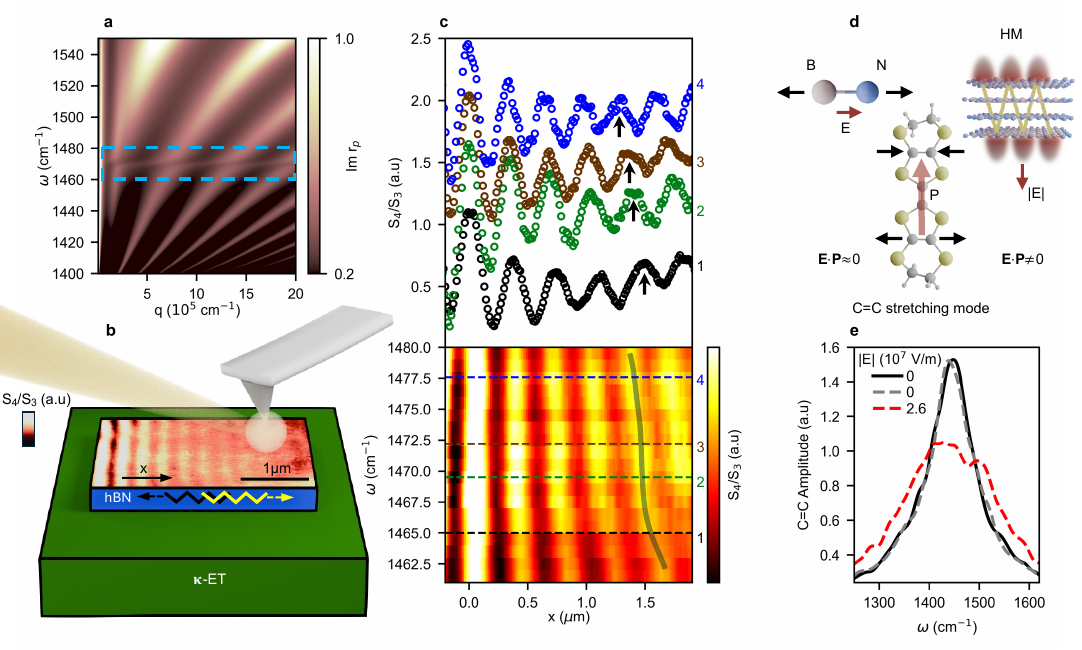}
    \caption{ \textbf{\boldmath Mode coupling at the hBN/{\ET} interface. a,} Phonon-polariton dispersion for an hBN/{\ET} interface (32-nm hBN, 100-nm {\ET}), represented by a false color map of the imaginary part of the Fresnel reflection coefficient $r_p$. Refer to Methods for calculation details. 
    \textbf{b,} s-SNOM experiment schematic on an hBN/{\ET} heterostructure, showing continuous-wave light backscattered by an atomic force microscope tip. Phonon-polaritons in hBN launched by the tip reflect off the hBN edge, forming interference patterns in the s-SNOM amplitude. The image shown was taken with incident light frequency $\omega = 1477.6$ cm$^{-1}$. The false color scale represents the normalized scattering amplitude S$_4$/S$_3$, where S$_4$ and S$_3$ are the near-field amplitude demodulated at the fourth and third harmonics of the tip-tapping frequency, respectively~\cite{S4S3}.  
    \textbf{c,} (Bottom) Hyperspectral image of S$_4$/S$_3$ at various illumination frequencies, where $x$ is along the phonon-polariton propagation direction. The overlay highlights the kink in the dispersion. Selected profiles are presented in the top panel. Black arrows mark the location of the fifth fringe. The profiles are vertically shifted for clarity.
    \textbf{d,} Schematic showing the coupling between the C=C stretching resonance and the hyperbolic mode of hBN enabled by the out-of-plane component of the  electric field of the HMs discussed in the main text and further analyzed in Supplementary Information Sections 2,3.
    \textbf{e,} 
    The calculated spectrum of the C=C stretching mode amplitude for BEDT-TTF molecules on hBN (black curve). The gray curve has phonon-phonon scattering suppressed. The red curve is the spectrum in the presence of an oscillating out-of-plane electric field from the HMs of hBN, with field strength $2.6\cdot 10^7$~V/m. Refer to Supplementary Information Section 11 for calculation details.
    }
     
    \label{fig:SNOM}
\end{figure*}

Next, we investigate resonant mode coupling across the hBN/{\ET} interface.
The hBN crystal hosts  HMs that exhibit light-like dispersion near the transverse optical (TO) frequency and phonon-like dispersion near the longitudinal optical (LO) frequency~\cite{basov2016}. Here, TO and LO refer specifically to the in-plane modes.
In practice, the HMs manifest as hyperbolic phonon-polaritons (HPhPs) that can be visualized in nano-optical experiments as originally done in Ref.~\cite{basov2104}.
We examine the HPhP dispersion of hBN upon interfacing with {\ET} through the calculated imaginary part of the p-polarized Fresnel reflection  $\operatorname{Im}r_p$, as shown in Fig.~\ref{fig:SNOM}a (Supplementary Information Section 10). For free-standing hBN, the HPhP dispersion displays a series of branches between the TO and LO phonon frequencies, each with a continuous dispersion~\cite{basov2104}. 
By contrast, for an hBN/{\ET} heterostructure, the HPhP branches are interrupted near the C=C stretching frequency (light blue box in Fig.~\ref{fig:SNOM}a).
Avoided crossings are visible, yet the gaps in the HPhP dispersion are not fully developed due to broadening caused by dielectric loss associated with the in-plane response of {\ET}~\cite{Sedlmeier,Faltermeier} (Supplementary Information Section 10). 

To experimentally examine the mode coupling at the hBN/{\ET} interface, we used scattering-type scanning near-field optical microscopy (s-SNOM)~\cite{hillenbrand2025visible}. Specifically, we evaluated the HPhP dispersion of hBN as its modes couple to the C=C stretching mode of {\ET}. We illuminated the metallized tip of an atomic force microscope with a frequency-tunable IR laser. The strong electric field with high momentum achieved near the tip apex launches polaritonic waves. We performed nano-IR measurements at $T=50$~K on a region of an hBN/{\ET} heterostructure with 32-nm hBN on 100-nm {\ET}. 
Fringes arising from the interference of propagating HPhPs were observed in the nano-IR images: representative results are shown in Fig.~\ref{fig:SNOM}b for $\omega = 1477.6$~\wavenum. Line profiles from 
s-SNOM collected with varying incident light frequency show that the HPhP wavelength changes with photon energy (Fig.~\ref{fig:SNOM}c). 
In free-standing hBN, the real-space fringes arising from HPhP interference evolve smoothly with the frequency~\cite{Bylinkin2021}. However, in our data on {\ET}-supported hBN, the fringe evolution exhibits an evident kink near the C=C frequency, indicated by the solid line tracking the fifth peak in Fig.~\ref{fig:SNOM}c.
These nano-IR data attest to mode coupling across the hBN-{\ET} interface

To gain insight into the experimentally observed mode coupling, we take a closer look at the interplay between the zero-point fluctuations of the hBN HMs and the C=C stretching mode. First-principles molecular Langevin dynamics simulations indicate that the coupling arises from the interaction of the transverse component of the electric field of the HMs with the C=C stretching mode. 
The C=C stretching mode exhibits a dipole moment that is directed primarily out-of-plane (Fig.~\ref{fig:SNOM}d and Supplementary Information Section 2).
It is important to note that the electric field of the HM has an out-of-plane component that can couple to the C=C mode (Fig.~\ref{fig:SNOM}d). The combined contributions of many HM branches 
reaching high momenta make this field extend far beyond the hBN surface (Supplementary Information Section 3).
We therefore calculate the C=C amplitude in the presence of an oscillating out-of-plane electric field with a frequency spectrum spanning the C=C frequency (fundamental frequency $\omega_0 = 1440$ cm$^{-1}$). Langevin molecular dynamics analysis carried out at $T=2$ K (Supplementary Information Section 11) shows that the out-of-plane electric field induced by the zero-point HM fluctuations reduces the C=C amplitude and splits the peak in Fig.~\ref{fig:SNOM}e. 
We therefore conclude that zero-point HM motion mediates the coupling between HMs
 and molecular vibrations without the need for external photons.

\subsection*{Hyperbolic cavity electrodynamics}

The data in Figs.~1-4 demonstrate that cavity electrodynamics plays a pivotal role in suppressing the superfluid density -- a hallmark equilibrium attribute of superconductivity.
These experiments also establish a precedent for successfully altering a material's ground state thermodynamic property through cavity engineering. Our specific implementation of the resonant structure exploits two notable features of hyperbolic cavity control. 
First, HMs entail both in-plane and out-of-plane components of the fluctuating electromagnetic fields. The out-of-plane component is crucial for coupling to the primarily out-of-plane dipole moment of the C=C stretching mode in the {\ET} superconductor under investigation. Second, the multiple branches of the HMs enable interaction with the weakly dispersing C=C resonance at multiple intersections over a broad range of momenta, as shown in  Fig.~\ref{fig:SNOM}a. Future experiments need to evaluate the spatial extent of the mode coupling away from the hyperbolic interface; our earlier data for a similar resonant hBN/MoO$_3$ heterostructure suggest that the relevant length scale may exceed $\sim 10^2$~nm~\cite{Sternbach2023}.

Our data and modeling suggest a plausible pathway for cavity enhancement of superconductivity. As an illustrative example, we consider the scenario of an optical phonon in a superconductor coupled to a hBN cavity within the framework of Bardeen, Cooper and Schrieffer (BCS) theory (Supplementary Information Section 13).
Electron-phonon coupling, quadratic in the phonon coordinate~\cite{sentef2018cavity,Kennes2017}, results in a suppression of the superfluid density, while linear coupling results in enhancement. 
This analysis suggests that conventional superconductors may exhibit increased superfluid density through interactions with HMs. Here we pause to caution the reader: {\ET} studied in our work is an unconventional superconductor~\cite{Cavalleri2021}, and BCS-derived inferences can only guide our intuition. A rigorous description of the physics involved in cavity-coupled {\ET} awaits further investigation. We conclude by highlighting that the concept of employing readily reconfigurable vdW hyperbolic cavities for resonant manipulation of material properties is rather general.
HMs of phononic, plasmonic and excitonic origin are hosted in many vdW materials including insulators, semiconductors as well as a variety of emerging materials~\cite{basov2025polaritonic}. These modes span the range from THz to visible~\cite{ruta2025good}, thus offering limitless options for hyper-cavity engineering.

\begin{footnotesize}
    
\renewcommand{\refname}{\vspace{-1em}}
\bibliographystyle{naturemag}

\bibliography{Bib}

\end{footnotesize}
\clearpage
\section*{Methods}

\subsection*{Device fabrication}

The heterostructures were fabricated by placing vdW materials of thicknesses on the order of 10 nm to 120 nm, on large flat surfaces of {\ET} single crystals (ac crystallographic plane, Supplementary Information Section 2). 
Measurements were performed on six devices (Supplementary Information Section 4).  Data in Fig.~\ref{fig:z-specs} are from Device 1, in
Fig.~\ref{fig:SNOM} from Device 2, in
Fig.~\ref{fig:imaging}(b-d) from Device 1, and in 
Fig.~\ref{fig:imaging}a from Device 6.
The data in Fig.~\ref{fig:Quantitative} represent a compilation of results from Devices 1, 3, 4, and 5.

The {\ET} crystal surfaces were either as-grown (Devices 1, 4, 5), exposed by cleaving bulk crystals (Devices 3 and 6), or exposed by exfoliating single crystals (Device 2). For bulk samples, the {\ET} single crystal was fixed onto a silicon substrate using either epoxy or silver paint. For Devices 3 and  6, the crystal was then cleaved to expose new surfaces. Large atomically flat surfaces were then selected by optical and atomic force microscopy. As-grown surfaces were regularly found to be atomically flat. Cleaved surfaces showed larger variations in surface roughness. For each device, MFM measurements spanned time periods ranging from weeks to months, and in general included multiple thermal cycles. For all devices, no sign of degradation was found.

Both hBN and {\RuCl}
microcrystals were exfoliated using a standard scotch-tape technique onto a polydimethylsiloxane (PDMS) stamp and identified via optical and atomic force microscopy. The stacking process was performed using a commercial transfer stage with temperature control. The PDMS stamp was gently pressed onto the {\ET} surface and then slowly retracted.
The hBN microcrystals were released at approximately 60°C, while the {\RuCl} microcrystals in Devices 1 and 5 were released at around 110°C and 130°C, respectively.
Isotopically pure hBN ($^{10}$BN) was used to achieve a narrow linewidth for the TO phonon. The narrow linewidth makes HPhPs visible in s-SNOM.

hBN/BSCCO samples were prepared by exfoliating and assembling BSCCO and hBN crystals in a glovebox filled with inert gas ($\mathrm{O_2}<0.1$ ppm; $\mathrm{H_{2}O<0.1}$ ppm). BSCCO was first exfoliated on SiO$_2$ substrates. hBN microcrystals were then exfoliated on a PDMS stamp and transferred onto BSCCO at room temperature.

\subsection*{s-SNOM}

The nano-infrared scattering experiments were performed using a home-built cryogenic scattering-type scanning near-field optical microscope (s-SNOM) housed in an ultra-high vacuum chamber with a base pressure of $\sim 7\times10^{-11}$ torr. The s-SNOM is based on a tapping-mode atomic force microscope (AFM) with the tapping frequency and amplitude of the AFM set to approximately 285~kHz and 70~nm, respectively. The s-SNOM operates by scattering tightly focused laser light from a sharp AFM tip, enabling near-field interactions to be probed with nanometer-scale resolution. The laser source is a tunable quantum cascade laser from Daylight Solutions. The laser beam was focused onto the metallized AFM tip using a parabolic mirror with a 12-mm focal length. The backscattered light was registered by a mercury cadmium telluride detector and demodulated following a pseudoheterodyne scheme. The signal was demodulated at the $n$th harmonic of the tip tapping frequency, yielding a background-free near-field signal. In this work, we selected $n$=3 and 4 to eliminate the far-field background.

\subsection*{MFM}

MFM measurements were performed in an Attocube cantilever-based cryogenic AFM (attoAFM I with attoLIQUID 2000 cryostat), where the microscope sits in a helium exchange gas
at low temperature. Nanosensors PPP-MFMR probes with hard magnetic coatings and force constants near $\mathrm{k}=2.8$ N/m were used for all measurements. 
Measurements were taken on different samples using different tips, all with resonance frequencies ranging from 75.5 kHz to 79.4 kHz. 
Measurements recorded the resonant frequency shift of the probe cantilever ($\mathrm{\Delta f}$) and were converted to $\dfdz$ via $\dfdz = - 2\mathrm{k}/\mathrm{f_0}\times\mathrm{\Delta f}$ ~\cite{Hartmann1999}, using the force constant $k$ provided by the manufacturer for each cantilever and the measured resonant frequency far from the sample surface, $f_0$.
The tip oscillation amplitude was typically 50 nm - 70 nm.  hBN is non-magnetic and therefore has no impact on the detection of the Meissner effect over hBN/{\ET}.  {\RuCl} is antiferromagnetic at low temperature, but has a low susceptibility.

To minimize noise and instrument drift, $\Delta f (z)$ was typically averaged over multiple measurements repeated at each point and taken with high $z$ resolution (typically 1-2~nm to define the surface accurately). During the measurements, the tip was electrically biased to compensate for the local contact potential difference and thus minimize electrostatic forces. The average curves were then smoothed with a uniform filter with width 30 to 50 pixels to reduce the appearance of noise.  For the {\ET} heterostructures, a linear background was subtracted by fitting the data far from the sample surface where the $\dfdz$ is negligible. For the BSCCO sample, a background measured on the silicon substrate was subtracted (Supplementary Information Section 6).
For proper comparison in Fig.~\ref{fig:z-specs}, we have compensated for the thickness of the hBN and {\RuCl} microcrystals such that the measurements for hBN/{\ET}, bare {\ET}, and {\RuCl}/{\ET} all have the {\ET} surface at zero.

Constant-height MFM images were taken with the tip at a fixed height above the plane of the \ET\ surface. The images shown are raw data, except for the linear conversion from $\Delta f$ to $\dfdz$ and an overall constant offset to calculate $\Delta\dfdz$. During the measurements, the tip was held at a fixed electric bias to compensate for the contact potential difference at one point in the field of view. 

The data points in Fig.~\ref{fig:Quantitative} were obtained either by fitting $\dfdz (z)$ curves or by converting constant-height MFM images to maps of the local $\lambda_\mathrm{in}$, the decay length for currents running parallel to the surface. $\lambda_\mathrm{in}$ was converted to superfluid density via the proportionality to $1/\lambda_\mathrm{in}^2 $, as described in Supplementary Information Section 5.

\subsection*{HPhP dispersion simulation}
The HPhP dispersion was calculated using the transfer matrix method. The dispersion is represented by peaks in the imaginary part of the Fresnel reflectivity of the device for p-polarized light, $r_p$. In the dispersion calculation, we consider the real-valued frequencies and momentum. For coupling analysis, the eigenmodes are found by determining the poles of $r_p$, where the complex frequencies and real momentum are considered. 

\subsection*{Data availability}
All datasets generated and analyzed during this study are available upon request from the authors.

\vspace{1em}

\begin{footnotesize}

\noindent{\footnotesize\textbf{Acknowledgments}} We acknowledge helpful discussions with O. Auslaender.
The authors acknowledge the use of facilities and instrumentation supported by NSF through the Columbia University, Columbia Nano Initiative, and the Materials Research Science and Engineering Center DMR-2011738.
D.N.B is a Moore Investigator in Quantum Materials EPIQS GBMF9455. Research at Columbia on electrodynamics of van der Waals interfaces is supported  by DOE-BES DE-SC0018426. Fabrication of vdW heterostructures at Columbia and the development of nano-optical methods is supported as part of Programmable Quantum Materials, an Energy Frontier Research Center funded by the U.S. Department of Energy (DOE), Office of Science, Basic Energy Sciences (BES), under award DE-SC0019443. Advances in MFM imaging at Columbia are supported by ARO award W911NF2510062.
This work was supported by the European Research Council (ERC-2024-SyG- 101167294 ; UnMySt), the Cluster of Excellence Advanced Imaging of Matter (AIM), Grupos Consolidados (IT1249-19) and SFB925,
We acknowledge support from the Max Planck-New York City Center for Non-Equilibrium Quantum Phenomena. The Flatiron Institute is a division of the Simons Foundation.
Support for hexagonal boron nitride crystal growth was provided by the Office of Naval Research, Award No. N00014-22-1-2582.
K.K. acknowledges the support of the JSPJ KANENHI (21K18144), the JSPS Core-to-Core Program (JPJSCCA20240001) and the Alexander von Humboldt Foundation.
The work at BNL was supported by the US Department of Energy,
office of Basic Energy Sciences, contract no. DOE-SC0012704.
D.M. and M.C. acknowledge support from the Gordon and Betty Moore Foundation’s EPiQS Initiative, Grant GBMF9069.
D.M.K. and M.A.S. acknowledge funding by the Deutsche Forschungsgemeinschaft (DFG, German Research Foundation) -  531215165 (Research Unit ‘OPTIMAL’).
M.A.S. was funded by the European Union (ERC, CAVMAT, project no. 101124492).
T.O. acknowledges support from the JSPS Overseas Research Fellowships.
D.S. was supported by the National Research Foundation of Korea (NRF) grant funded by the Korea government (MSIT) (No. RS-2024-00333664). J. Y. and Q.L were supported by the Office of Basic Energy Sciences, Materials Sciences and Engineering Division, U.S. Department of Energy (DOE) under Contract No. DE-SC0012704. 
We acknowledge project/application support by the Max Planck Computing and Data Facility.

\vspace{1em}

\noindent{\footnotesize\textbf{Author contributions}}
I.K., T.A.W., J.X., and G.P. collected and analyzed MFM data. I.K. and T.A.W. wrote the manuscript. S.Z. and S.S.Z. collected SNOM measurements. S.Z and J.Z. analyzed SNOM data. S.Z. simulated the HPhP dispersion for hBN on \ET. D.S. and B.S.Y.K. fabricated the heterostructures with input from C.R.D. G.P. performed SEM imaging of the MFM tips. J.Y. and Q.L. performed SQUID magnetometry measurements. T.O. provided simulations of the electric field in a z-dependent superfluid density scenario, and in an oscillating dipole above hBN scenario. J.H.E. grew the hBN. M.J., R.P., and S.W. provided discussion and interpretation of the data, as well as formulation of the idea. K.M. and K.K. grew the {\ET} crystals. G.G grew the BSCCO crystal. D.M. and M.C. grew the {\RuCl} crystal. M.B. and A.C. provided discussion and interpretation of the data. D.M.K. provided guidance on fitting and participated in the discussion on mechanisms.  M.H.M. computed the quantum electric field fluctuations from a thin film of hBN. D.S. performed the ab initio calculations under the supervision of A.R. D.S., M.H.M., and A.R. analyzed the ab initio results and developed a microscopic model for the quenching of superconductivity. E.V.B. developed the analytical model for BCS-HM coupling. M.A.S. provided theoretical input on the interpretation of results and contributed to the manuscript. A.N.P. advised on measurements and data interpretation. D.N.B. advised and supervised the project. All authors discussed the results and contributed to the final paper.

\vspace{1em}

\noindent{\footnotesize\textbf{Competing interests}} The authors declare no competing interests.

\vspace{1em}

\noindent{\footnotesize\textbf{Correspondence and requests for materials}} should be addressed to Itai Keren, Tatiana Webb, Shuai Zhang or Dmitri Basov.

\end{footnotesize}


\clearpage

\twocolumn

\end{document}

%% file: Preamble.tex
\DeclareUnicodeCharacter{0308}{\"{}}
\usepackage[a4paper,top=2cm,bottom=2cm,left=1cm,right=1cm,marginparwidth=1.75cm]{geometry}
\usepackage{graphicx} 
\usepackage{amsmath}
\usepackage{comment}
\usepackage[numbers,sort&compress]{natbib}
\usepackage{color}
\usepackage{braket}
\usepackage[usenames, dvipsnames]{xcolor}
\usepackage{bm}
\usepackage{hyperref}
\hypersetup{
    colorlinks=true,
    linkcolor=blue,     
    urlcolor=blue,
    citecolor=blue
}

\usepackage{authblk}
\usepackage{marvosym}  
\usepackage[font=small]{caption} 
\usepackage[normalem]{ulem}

\newcommand{\mainpoint}{}

\newcommand{\dfdz}{\partial_{z} F_{z}}
\newcommand{\Deltadfdz}{\Delta\partial_{z} F_{z} }

\newcommand{\RuCl}{RuCl$_3$}
\newcommand{\ET}{$\kappa$-ET}
\newcommand{\Tc}{T_{\textrm{c}}}

\newcommand{\wavenum}{cm$^{-1}$}
\newcommand{\rhoeff}{\rho_{\textrm{eff}}}

\renewcommand{\thefootnote}{\fnsymbol{footnote}}